\documentclass[
 reprint,
 amsmath,amssymb,
 aps,
prb,
showkeys
]{revtex4-2}

\usepackage{graphicx}
\usepackage{subfigure}
\usepackage{dcolumn}
\usepackage{bm}
\usepackage{color}
\usepackage{hyperref}
\usepackage{siunitx}
\usepackage{booktabs}
\usepackage{multirow}
\usepackage{makecell}
\usepackage[none]{hyphenat}
\usepackage{newunicodechar}
\newunicodechar{★}{\ensuremath{\bigstar}}
\begin{document}
\preprint{APS/123-QED}

\title{On the Nonlinear Dependence of Underground Muon Rate on Atmospheric Temperature Observed at Daya Bay}

\author{Lei Liao}
\author{Taichong Ge}
\author{Zhe Wang}\email{Correspondence: wangzhe-hep@mail.tsinghua.edu.cn}
\affiliation{%
Department of Engineering Physics, Tsinghua University, Beijing 100084, China \\
Center for High Energy Physics, Tsinghua University, Beijing 100084, China\\
Key Laboratory of Particle \& Radiation Imaging (Tsinghua University), Ministry of Education, Beijing 100084, China
}

\date{\today}

\begin{abstract}
The underground cosmic-ray muon rate is seasonally modulated by atmospheric temperature. It can be explained by the theories of Barrett, Gaisser, and others. However, the Daya Bay Neutrino Experiment observes a nonlinear temperature dependence.
When deriving the temperature dependence of the muon rate, existing theories consider only the impact of the local temperature on muon production at the production layer. 
In this work, we use a more general solution to the cascade equations that fully depicts how the entire temperature profile influences the final muon rate.
The corresponding new definitions of the effective temperature weight and temperature dependence coefficient are also presented.
We have examined the results using the numerical tool MCEq with real atmospheric temperature inputs. A linear modulation is recovered and verified. 
This work can help explain the nonlinear effect observed at Daya Bay and provide a more refined theoretical framework for calculating the temperature coefficient in other experiments.
\end{abstract}

\maketitle

\section{Introduction}
The modulation of the underground cosmic-ray muon rate by atmospheric temperature has been observed in early underground experiments, such as~\cite{Barrett,Sherman,Poatina, Baksan, MACRO, AMANDA, IceCube, MINOS, Borexino, GERDA, DoubleChooz}. 
The single muon event rate is positively correlated with atmospheric temperature changes, as explained by the theories of Barrett et al~\cite{Barrett}, Gaisser~\cite{Gaisser}, MACRO Collaboration~\cite{MACRO}, and Grashorn et al.~\cite{KPionRatio}.
However, at the Daya Bay Neutrino Experiment, a nonlinear behavior is observed in studies of the modulation effect in the shallower Experimental Halls 1 and 2~\cite{DYB-ICRC2025, DYB-TAUP2025}. 
The correlation appears stronger for warmer months than for cooler months. 
Further studies are in progress~\cite{Bangzheng}, and no conclusion has yet been reached.

In the solutions for the cosmic-ray cascade evolution equations, we found that the meson or muon energy spectra in the existing theory~\cite{Gaisser} are obtained by simply linking approximate solutions valid in the high- or low-energy limits. 
These approximate solutions are derived under the assumption of an isothermal atmosphere. 
In the work of Grashorn on effective-temperature derivation~\cite{KPionRatio}, the depth-dependent response to an arbitrary temperature perturbation is achieved after the initial isothermal solution. Therefore, the conventional $T_{\rm eff}$, $\Delta T_{\rm eff}$, and $\alpha_T$ should be understood as isothermal projected quantities rather than direct functional derivatives with respect to a realistic non-isothermal temperature profile.

In this work, in Sec.~\ref{sec:solution}, we start with an improved solution of the cascade evolution equations for the meson and muon energy spectra, which naturally account for the real and complex atmosphere temperature profile and cover both high- energy and low-energy regimes.
After the preparation of this work, we found 
In Sec.~\ref{sec:weight}, new definitions of \(T_{\rm eff}\), \(\Delta T_{\rm eff}\), and \(\alpha_T\) are provided by taking the functional derivative of the muon rate with respect to the actual temperature profile into account.
To test our new prediction, we use MCEq~\cite{MCEq_source}, a numerical tool for atmospheric shower cascade equations, to simulate atmospheric showers (Sec.~\ref{sec:MCEq}). 
With our new solution, the underground muon rate recovers a linear dependence on atmospheric temperature. The work will be summarized in Sec.~\ref{sec:summary}.

\section{A more general solution to shower development equations}
\label{sec:solution}

The solution for the underground muon rate is derived in this section. Its dependency on the complete air temperature profile is preserved. 
After independently preparing this manuscript, we found a similar study in Ref.~\cite{Previous}. However, this section is still useful for understanding the procedure of this paper, and after this section, this work and \cite{Previous} are different.

\begin{figure}[h]
  \centering  \includegraphics[width=0.4\textwidth]{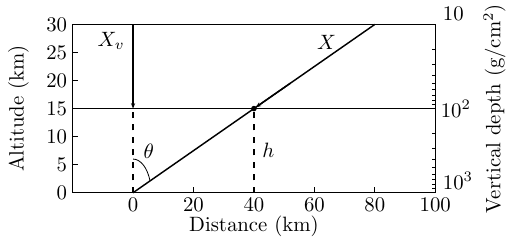}
  \caption{Slant depth $X$, its vertical component, $X_v$, local vertical altitude, $h$, and the zenith angle, $\theta$. The convention is consistent with Ref.~\cite{Gaisser}.}
  \label{fig:slant_depth}
\end{figure}

We follow the convention in Gaisser~\cite{Gaisser} about the slant depth, $X$, the local vertical altitude, $h$, and the zenith angle, $\theta$. The vertical component of $X$ is $X_v$, denoted with a subscript of $v$, with $X_v=X\cos\theta$. 
Their relations can be seen in Fig.~\ref{fig:slant_depth}.

Please note that for a function, for example, $d(E,X,T)$, we sometimes do not print certain parameters, but when they are important for the current discussion, we will include them again. The function $d(E,X,T)$ can be shown as $d$, $d(E,X)$, or $d(E,X,T)$. Hope this doesn't cause any confusion. 

\subsection{Meson energy spectrum at air depth $X$}

The temperature response is generated primarily by charged pions and kaons, collectively denoted as $M = \pi$ or $K$. 
The competition between their decay and interaction results in a dependence on air temperature.
Prompt particles are omitted because their decay lengths are too short for the atmospheric interaction-decay competition to produce a comparable density-profile response. 
The cascade is treated in one dimension along the slant-depth coordinate $X$, neglecting transverse shower development.

Following the meson cascade equation in Chapter 5 of Gaisser~\cite{Gaisser}, the MCEq paper~\cite{MCEq_source}, and Ref.~\cite{KPionRatio}, the differential spectrum of a conventional meson satisfies
\begin{equation}
\begin{split}
    \frac{d\phi_M (E,X)}{dX} &=
        - \frac{\phi_M (E,X)}{d_M (E,X)}
        - \frac{\phi_M (E,X)}{\Lambda_M}\\
        &+ \frac{Z_{NM}}{\lambda_N} \phi_N(E) e^{-X/\Lambda_N}.
\end{split}
\end{equation}
Here $\phi_M (E,X)$ is the meson differential energy spectrum at slant depth $X$ and $E$ is the meson energy. Its change at the level of $X$ to $X+dX$ depends on the meson decay and interaction, and generation from nucleons, and these contributions are the first, second, and third terms, respectively, on the right-hand side of the equation.
$\phi_N (E) \rm exp(-X/ \Lambda_N)$ is the attenuated nucleon spectrum at depth $X$. 
$\lambda_N$ is the nucleon interaction length, and $Z_{NM}$ is the spectrum-weighted moment, which characterizes the meson production by a spectrum of nucleons,
The strength $Z_{NM}/\lambda_N$ follows from the scaling approximation and a power-law primary spectrum, which allows inclusive meson production to be represented by the spectrum-weighted moment $Z_{NM}$.
$d_M$ is the meson decay length. $\Lambda_M$ is the meson attenuation length, where $\Lambda_M = \lambda_M/(1-Z_{MM})$ and $\lambda_M$ and $Z_{MM}$ are the meson interaction length and the spectrum-weighted moment for meson-to-meson. More detailed information about them can all be found in Ref.~\cite{Gaisser}.

The hadronic interaction lengths in $\rm{g/cm^2}$ are independent of the local density. For instance, the meson interaction length is 
\begin{equation}
    \lambda_M=\frac{Am_N}{\sigma_{M}},
\end{equation}
in which $A$ is the mean mass number of atmospheric molecules, $m_N$ is the average nucleon mass, and $\sigma_{M}$ is the average interaction cross-section for the mesons.
The only coefficient having an explicit functional of the atmospheric profile is the decay length
\begin{equation}
    d_M (E,X) = \rho(X,T) \gamma c \tau_{M}
              = \frac{c\tau_ME}{m_M c^2}\rho(X,T),
    \label{eq:OmitT}
\end{equation}
where $\gamma$ is the Lorentz factor, $c$ is the speed of light, E is the meson energy, and $\tau_M$ and $m_M$ are the meson lifetime and mass, respectively.
$\rho(X,T)$ is a depth- and temperature-dependent air density. 

With the boundary condition $\phi_M(E,0)=0$, the solution by the integrating-factor method is
\begin{equation}
\begin{split}
    \phi_M(E,X) &= 
        \frac{Z_{NM}}{\lambda_N} \phi_N (E)
        \int^X_0 \exp [ 
            -\frac{X'}{\Lambda_N}
            -\frac{X-X'}{\Lambda_M} \\
            &-\int^X_{X'} \frac{dX''}{d_M (E,X'')} 
        ] dX'.
    \label{eq:FirstNew}
\end{split}
\end{equation}
The outer integral accounts for all possible meson production depths $X'$.
The first exponential factor gives the attenuation of the parent nucleon flux before production, and the second describes meson attenuation by hadronic interactions after production.
The innermost integral is the cumulative decay probability along the meson trajectory.
Using the expression for the decay length, the innermost decay kernel can be written as
\begin{equation}
    \int^X_{X'} \frac{dX''}{d_M (E,X'')} = 
    \frac{m_M c^2}{c\tau_M E} \int_{X'}^X \frac{dX''}{\rho(X'',T)},
\end{equation}

The resulting expression of the meson energy spectrum at slant depth $X$ in Eq.~\ref{eq:FirstNew} has a nonlocal feature because the meson flux at depth $X$ depends on the density along every interval $X' < X'' < X$ crossed by the meson surviving probability.
Introducing the double-integral form keeps the production-depth integral and the decay-history integral distinct, and no isothermal atmosphere is assumed.
This is the key information required for a functional temperature response.

\subsection{Muon production energy spectrum at air depth $X$}

The next step is to convert the meson spectrum into a local muon production source.
The production term $Q_{\mu}(E_\mu,X)$ is defined as the number of muons produced per unit slant depth and per unit muon energy at depth $X$ from the decay of a parent meson.

For a parent meson with $E$, the number of decays in a layer $dX$ is proportional to $\phi_M / d_M$.
The corresponding contribution to muons of energy $E_{\mu}$ is obtained by folding the meson decay rate with the muon production distribution $dn_{\mu}/dE_{\mu}$, in the same form used in Gaisser's derivation~\cite{Gaisser} of the atmospheric muon spectrum
\begin{equation}
     Q_{\mu}(E_\mu,X)=
        \int dE
        \frac{\phi_M(E, X)}{d_M(E,X)}
        \frac{dn_{\mu}(E_{\mu}, E)}{dE_{\mu}}.
    \label{eq:2bodydecay}
\end{equation}

For the two-body decay $M^\pm\rightarrow\mu^\pm+\nu_\mu(\bar{\nu}_\mu)$, a muon of energy $E_{\mu}$ can be produced only by a parent in the energy interval $E_{\mu} \leq E \leq E_{\mu}/r_M$, where $r_M = m^2_{\mu} / m^2_M$.
In the laboratory frame, the decay distribution is flat within this interval, $dn_{\mu}/dE_{\mu} = B_M [(1-r_M)E]^{-1}$, in which $B_M$ is the branching ratio of $M\to \mu$. 
The production spectrum generated by mesons of meson $M$ is therefore
\begin{equation}
    Q_{\mu}(E_\mu,X)=
        \frac{m_M c^2 B_M}{c \tau_M \rho(X,T)(1-r_M)}
        \int_{E_\mu}^{\frac{E_\mu}{r_M}}
        \frac{dE}{E^2}
        \phi_M(E,X).
    \label{eq:QmuBR}
\end{equation}

The general meson spectrum with a non-isothermal air temperature profile in Eq.~\ref{eq:FirstNew} is introduced, and we obtain
\begin{equation}
\begin{split}
    Q_{\mu}&(E_\mu,X) \\
        &=
        \frac{Z_{NM} m_M c^2 B_M}{\lambda_N c\tau_M \rho(X,T)(1-r_M)}\\
        &\cdot\int_0^X dX'\,
        \mathrm{exp} \left[
            \frac{-X'}{\Lambda_N}
            -\frac{X-X'}{\Lambda_M}
            \right] \\
        &\cdot\int_{E_\mu}^{E_\mu/r_M}
        \frac{dE}{E^2}\,
        \phi_N(E)
        \mathrm{exp} \left[
            -\frac{m_M c^2}{c\tau_M E} \int^X_{X'} \frac{dX''}{\rho(X'',T)}
            \right].
            \label{eq:MX}
\end{split}
\end{equation}
This expression combines three ingredients or integrals: the meson two-body decay kinematic interval in energy, the possible production depth of the parent mesons, and the production depth of the grandparent nucleons.

Equation~\ref{eq:MX} shows an explicit density dependency at two places.  
At a certain slant depth interval $X$ to $X+dX$, a small local density will enhance the fraction of meson decay and introduce more muons; see Eq.~\ref{eq:2bodydecay}. But along the path of mesons, $X'$ to $X$, a smaller density will also cause fewer mesons to survive and reach the slant depth $X$ with a high energy, and a smaller density path will make a smaller muon contribution at the interval $dX$; see Eq~\ref{eq:FirstNew}. The roles of density could be contradictory. 
The functional dependency needs to be retained thoroughly.

\subsection{Energy loss and muon decay and muon energy spectrum on the ground}

We will work out the muon energy spectrum at slant depth $X$ from the decay of a type of meson. The energy spectrum is assumed to be $\phi_{\mu}(E_\mu,X)$.

The energy loss of a muon after passing an interval $dX$ is~\cite{Gaisser} 
\begin{equation}
    \frac{dE_\mu}{dX}\approx-\alpha=-2~\rm{MeV/(g~cm^{-2})}.
\end{equation}
The change of $\phi_{\mu}(E_\mu,X)$ in $dX$ is 
\begin{equation}
\frac{\partial \phi_{\mu}}{\partial X}=\alpha \frac{\partial \phi_{\mu}}{\partial E_\mu}.
\end{equation}

The change of $\phi_\mu$ after considering the decay of muon in an interval $dX$ is
\begin{equation}
    \frac{d\phi_{\mu}}{dX}=-\frac{m_\mu c^2}{c\tau_\mu E_\mu \rho(X,T)}\phi_\mu.
\end{equation}

Taking both the muon energy loss and decay effects into account, the partial differential equation for $\phi_\mu(E_\mu, X)$ is
\begin{equation}
    \frac{\partial \phi_{\mu}}{\partial X}=-\frac{m_\mu c^2}{c\tau_\mu E_\mu \rho(X,T)}\phi_{\mu} + \alpha \frac{\partial \phi_{\mu}}{\partial E_\mu} + Q_{\mu}(E_\mu,X).
\label{eq:pde_eloss_decay}
\end{equation}

The partial differential Eq.~\ref{eq:pde_eloss_decay} is solved using the method of characteristics. The solution is
\begin{equation}
\begin{split}
    \phi_{\mu}(E_\mu, X_{grd})&=\int_0^{X_{grd}} 
    Q_{\mu}(E_\mu+\alpha(X_{grd}-X), X)\\ 
    &\cdot P(E_\mu, X, X_{grd}) dX,
\label{eq:eloss_decay}
\end{split}
\end{equation}
in which $X_{grd}$ is the ground level $X$ and 
\begin{equation}
\begin{split}
    P&(E_\mu, X, X_{grd})=\\
    &\exp \left\{-\int_X^{X_{grd}} 
    \frac{m_\mu c^2}{c\tau_\mu \rho(\widetilde{X},T)}
    \frac{d\widetilde{X}}{[E_\mu+\alpha(X_{grd}-\widetilde{X})]}\right\}.
    \label{eq:P_eloss_decay}
\end{split}
\end{equation}
The integral in Eq.~\ref{eq:eloss_decay} means that at the ground level, the contribution from all slant depths, $Q_\mu$, is integrated, and a certain energy shift $\alpha(X_{grd}-X)$ is corrected. The second term, $P$, can be understood as the survival probability of the muons. $\phi_{\mu}(E_\mu, X_{grd})$ is the basic expression used next.

\subsection{Temperature and density of the atmosphere}

We assume the atmosphere has an onion-like structure. The density and temperature are constant in the horizontal (Earth's surface curvature) direction.
The density-temperature-pressure, $\rho$-$T$-$p$, relation used in the Gaisser~\cite{Gaisser} cascade calculation is retained here. 
Using the vertical height coordinate $h$, the ideal gas law and hydrostatic equilibrium can be written as
\begin{equation}
    \rho(h)=\frac{M_{\rm air}p(h)}{R_{\rm air}T(h)}
    \qquad{\rm{and}}\qquad
    \frac{dp(h)}{dh}=-\rho(h)g,
    \label{eq:idealGas}
\end{equation}
where $R_{\rm air}$ denotes the gas constant, $M_{\rm air}$ is the mean molar mass of air, and $g$ is the acceleration due to gravity.
The relation between pressure and vertical depth is 
\begin{equation}
\label{hydrostatic_eq}
    p(X_v)=
        gX_v.
\end{equation}
Combining this relation with the ideal gas law of Eq.~\ref{eq:idealGas} gives
\begin{equation}
    \rho(X,T)=
        \frac{gM_{\rm air}}{R_{\rm air}}\frac{X_v}{T(X_v)}.
        \label{eq:TX}
\end{equation}
With this simplified onion-like atmosphere, at a given slant depth of $X$, the density is determined algebraically by the temperature at the same vertical depth of $X_v$.

\subsection{Functional dependence of ground level muon spectrum on air temperature}

Substituting the relation derived in Eq.~\ref{eq:TX} into the triple-integral muon production energy spectrum in Eq.~\ref{eq:MX} gives
\begin{equation}
\begin{split}
    Q_{\mu}&(E_{\mu},X,\theta) 
        = 
        \frac{R_{air} Z_{NM} m_M c^2 B_M}{g M_{air} \lambda_N c{\tau}_M (1-r_M)}
        \frac{T(X_v)}{X_v}\\
        &\cdot\int^X_0 dX' \mathrm{exp} \left[
            -\frac{X'}{\Lambda_N}
            -\frac{X-X'}{\Lambda_M}
        \right] \\
        &\cdot\int^{E_{\mu}/r_M}_{E_{\mu}}
        \frac{dE}{E^2}
        \phi_N(E)\\
        &\cdot\mathrm{exp} \left[
            -\frac{R_{air} m_M c^2 }{g M_{air} c{\tau}_M E}
            \int^X_{X'} \frac{T(X''_v)}{X''_v} dX''
        \right].
        \label{eq:MuonProduction}
\end{split}
\end{equation}
Substituting the relation derived in Eq.~\ref{eq:TX} into the muon survival probability in Eq.~\ref{eq:P_eloss_decay} gives
\begin{equation}
\begin{split}
    P&(E_\mu, X, X_{grd},\theta) \\    
    &=\exp \{ -\int_X^{X_{grd}} 
    \frac{m_\mu c^2 R_{\rm air}}{c\tau_\mu gM_{\rm air}} \frac{T(\widetilde{X}_v)}{\widetilde{X}_v} \\
    &\cdot\frac{d\widetilde{X}}{[E_\mu+\alpha(X_{grd}-\widetilde{X})]} \}.
    \label{eq:PElossDecay}
\end{split}
\end{equation} 
The functional dependence on the temperature profile $T(Y_v)$ is explicitly seen in $Q_\mu$ and $P$.

\subsection{Underground muon rate}
\label{sec:MuonRate}
The underground muon rate is obtained by integrating over all muons above an energy threshold $E_{th}$ and over the full production solid angle. The energy threshold $E_{th}$ corresponds to the overburden at a specific muon incidence angle in an underground laboratory. The muon rate is
\begin{equation}
\begin{split}
    R_{\mu} 
        &=\int \sin\theta d\theta d\phi \int_{E_{th}(\theta,\phi)}^\infty dE_{\mu} \phi_\mu(E_\mu, X_{grd}(\theta,\phi),\theta).
\end{split}
\label{eq:MuonRate}
\end{equation}

Equations~\ref{eq:MuonRate} together with Eq.~\ref{eq:eloss_decay}, \ref{eq:MuonProduction}, and \ref{eq:PElossDecay} give the underground muon rate prediction.
\begin{equation}
\begin{split}
\phi_\mu&(E_\mu, X_{grd}(\theta,\phi),\theta) = \int_0^{X_{grd}(\theta,\phi)} \\
        &Q_{\mu}(E_\mu+\alpha(X_{grd}-X), X) 
        P(E_\mu, X, X_{grd}) dX, \\
Q_{\mu}&(E_\mu+\alpha(X_{grd}-X), X,\theta) 
    =\frac{T(X_v)}{X_v}
        \int^X_0 A dX' \\
        &\int^{[E_\mu+\alpha(X_{grd}-X)]/r_M}_{E_\mu+\alpha(X_{grd}-X)} B
        \mathrm{exp} \left[
            -C \int^X_{X'} \frac{T(X''_v)}{X''_v} dX''
        \right] dE,\\
P&(E_\mu, X, X_{grd},\theta)
    =\mathrm{exp} \left[-\int_X^{X_{grd}} 
             D \frac{T(\widetilde{X}_v)}{\widetilde{X}_v} d\widetilde{X}
        \right].\\
\end{split}
\label{eq:MuonRateSupp}
\end{equation}
The meanings of $A$, $B$, $C$, and $D$ can be easily determined by comparing them to their original expressions.

The muon production energy spectrum contains the atmospheric profile in three places.
The factor outside the depth integral of $Q_\mu$ is evaluated at the actual muon production depth $X$.
It refers only to the local slant-depth layer from $X$ to $X+dX$, where the parent meson decays and produces the muon. 
Higher temperature and lower density increase the fraction of meson decay and produce more muons; see Eq.~\ref{eq:2bodydecay}.

The exponential integral kernel of $Q_\mu$ contains the temperature history between the meson production depth $X'$ to the decay depth $X$.
The integral over $X''$ is nonlocal and sums over the possible production depths of that parent meson.
Along the path of $X'$ to $X$, higher temperature and smaller density will also introduce fewer mesons to reach the local depth level of $dX$; see Eq~\ref{eq:FirstNew}.

The third temperature factor in the survival probability $P$ describes the competition between muon decay and energy-loss processes. Along the muon propagation path from its production at $X$ to the destination of $X_{grd}$, higher temperature and lower density will also cause a smaller fraction of muons to reach the ground; see Eq.~\ref{eq:P_eloss_decay}.  

The nucleon hadronic attenuation factors remain independent of the temperature profile, while the two-body decay kinematics fixes only the limits of the energy integral.
This decomposition does not imply that the conventional effective-temperature weight is a purely local production-layer factor; rather, it distinguishes the local decay contribution from the propagation-history contribution prior to the isothermal projection used in the standard analytic treatment.

The pion and kaon contributions have the same structure.
Their contributions in Eq.~\ref{eq:MX} and \ref{eq:MuonRateSupp} can be added using their respective meson-decay parameters. 

\section{A new definition of effective temperature and linear dependence coefficient}
\label{sec:weight}

\subsection{$T_{\rm eff}$, $\Delta T_{\rm eff}$, and $\alpha_T$} 
Since we use a simple, horizontally uniform, onion-like atmosphere, the underground muon rate in Eq.~\ref{eq:MuonRate} is a functional of the temperature profile $T(Y_v)$. 
We therefore define the effective-temperature weight and the temperature coefficient from the first-order functional response around the actual mean atmospheric profile, rather than from an isothermal projected response.
Let $T_0(Y_v)$ be the time-averaged temperature profile at a given location, and let $\Delta T(Y_v)$ be the deviation from that average at a given time. Then the muon rate can be expanded to first order around $T_0(Y_v)$ as
\begin{equation}
\begin{split}
&R_{\mu}(T(Y_v)) \\
=& R_{\mu}(T_0(Y_v)+\Delta T(Y_v))\\
=& R_{\mu}(T_0(Y_v)) +\int_0^{X_{grd,v}} \frac{\delta R_{\mu}}{\delta T_0(Y_v)} \Delta T(Y_v) dY_v\\
=&R_{\mu 0}+ \Delta R_\mu,
\end{split}
\end{equation}
where $\frac{\delta R_{\mu}}{\delta T_0(Y_v)}$ is the functional derivative at $T_0(Y_v)$. The quantities $R_{\mu 0}$ and $\Delta R_\mu$ are the leading term and the first-order correction in the expansion, and they are the average muon rate and its variation at that time, respectively.

Following the convention used by the MACRO experiment~\cite{MACRO}, 
we redefined $T_{\rm eff}$, 
$\Delta T_{\rm eff}$, and $\alpha_T$ as follows
\begin{equation}
\begin{split}
T_{\rm eff} 
&= \frac {\int_0^{X_{grd,v}} \frac{\delta R_{\mu}}{\delta T_0(Y_v)} T_0(Y_v) dY_v} 
{\int_0^{X_{grd,v}} \frac{\delta R_{\mu}}{\delta T_0(Y_v)} dY_v},
\end{split}
\end{equation}

\begin{equation}
\begin{split}
\Delta T_{\rm eff} 
&= \frac{\int_0^{X_{grd,v}} \frac{\delta R_{\mu}}{\delta T_0(Y_v)} \Delta T_0(Y_v) dY_v}
{\int_0^{X_{grdv}} \frac{\delta R_{\mu}}{\delta T_0(Y_v)} dY_v},
\end{split}
\end{equation}

\begin{equation}
\begin{split}
\alpha_T
&= \frac{\int_0^{X_{grd,v}} \frac{\delta R_{\mu}}{\delta T_0(Y_v)} T_0(Y_v) dY_v} {R_{\mu 0}}.
\end{split}
\label{eq:alpha}
\end{equation}
With these definitions, the relative muon-rate variation depends linearly on the relative variation of the effective temperature:
\begin{equation}
\begin{split}
\frac{\Delta R_\mu}{R_{\mu 0}} &= \alpha_T \frac{\Delta T_{eff}}{T_{eff}}.
\end{split}
\end{equation}

\subsection{Weight and functional derivative}
Aligned with the derivation of Barrett~\cite{Barrett}, we redefine the effective temperature weight as the functional derivative of the counting rate with respect to the temperature profile
\begin{equation}
    W(Y_v) \equiv \frac{\delta R_{\mu}}{\delta T_0(Y_v)}.
\end{equation}
Note that the functional derivative is also a functional, not a function, and its interaction with a given function yields a scalar.
For example, the functional derivative of the counting rate with respect to the temperature curve, when combined with the distribution of changes in the temperature profile, gives the change in the counting rate.

\begin{widetext}
The functional derivative is
\begin{equation}
\begin{split}
\frac{\delta R_{\mu}}{\delta T_0(Y_v)}
    &=\int \sin\theta d\theta d\phi \int_{E_{th}(\theta,\phi)}^\infty dE_{\mu} 
    \int_0^{X_{grd}(\theta,\phi)} 
        \frac{\delta [Q_{\mu}(E_\mu+\alpha(X_{grd}-X), X) 
        P(E_\mu, X, X_{grd})]}{\delta T_0(Y_v)} dX,
\end{split}
\label{eq:Deri1}
\end{equation}
in which
\begin{equation}
\begin{split}
    &\frac{\delta [Q_{\mu}(E_\mu+\alpha(X_{grd}-X), X) P(E_\mu, X, X_{grd})]}{\delta T_0(Y_v)} \\
    &=\frac{\delta Q_{\mu}(E_\mu+\alpha(X_{grd}-X), X)}{\delta T_0(Y_v)} P(E_\mu, X, X_{grd})
    +Q_{\mu}(E_\mu+\alpha(X_{grd}-X), X)\frac{\delta P(E_\mu, X, X_{grd})}{\delta T_0(Y_v)}. 
\end{split}
\label{eq:Deri2}
\end{equation}

The key step is to calculate the functional derivative of the muon production spectrum $Q_\mu$ and the survival probability $P$ in Eq.~\ref{eq:MuonRate} and \ref{eq:MuonRateSupp}.
By introducing a functional variation of $\Delta T(Y_v)$ to the average $T_0(Y_v)$, the variation, $\delta Q_{\mu}$, on the $Q_\mu$ can be calculated.
\begin{equation}
\begin{split}
    &\delta Q_{\mu} = Q_{\mu} (T_0(Y_v)+\Delta T(Y_v))-Q_{\mu}(T_0(Y_v)) \\
    &= \int_0^\infty \delta(X_v-Y_v)
    \frac{\Delta T(Y_v)}{Y_v}dY_v
        \int^X_0 A dX'
        \int^{[E_\mu+\alpha(X_{grd}-X)]/r_M}_{E_\mu+\alpha(X_{grd}-X)} B
        \mathrm{exp} \left[
            -C\int^X_{X'} \frac{T_0(X''_v)}{X''_v} dX''
        \right]dE\\
    &- \frac{T(X_v)}{X_v}
        \int^X_0 A dX'
        \int^{[E_\mu+\alpha(X_{grd}-X)]/r_M}_{E_\mu+\alpha(X_{grd}-X)} B
        \mathrm{exp} \left[
            -C\int^X_{X'} \frac{T_0(X''_v)}{X''_v} dX''
        \right]dE
        C\int^{X_v}_{X'_v} \frac{\Delta T(Y_v)}{Y_v} \frac{dY_v}{\cos\theta}.
\end{split}
\end{equation}
The last integral over $dY_v$ above can be expressed as 
\begin{equation}
\begin{split}
\int^\infty_{0} \frac{\Delta T(Y_v)}{Y_v} 
\Theta(Y_v-X'_v) \Theta(X_v-Y_v) \frac{dY_v}{\cos\theta},
\end{split}
\end{equation}
where $\Theta$ is a step function, equal to 1 when its argument is positive and 0 otherwise. 
The integration sequence of $X'$ and $Y$ is exchanged. 
The functional derivative at $T_0(Y_v)$ is 
\begin{equation}
\begin{split}
&\frac{\delta Q_{\mu}}{\delta T_0(Y_v)} = [\frac{\delta Q_{\mu}}{\delta T_0(Y_v)}]_1 +
[\frac{\delta Q_{\mu}}{\delta T_0(Y_v)}]_2 \\   
&= \delta(X_v-Y_v)
    \frac{1}{Y_v}
        \int^X_0 A dX'
        \int^{[E_\mu+\alpha(X_{grd}-X)]/r_M}_{E_\mu+\alpha(X_{grd}-X)} B
        \mathrm{exp} \left[
            -C\int^X_{X'} \frac{T_0(X''_v)}{X''_v} dX''
        \right]dE\\
&-\frac{T(X_v)}{X_v Y_v \cos\theta}  \Theta(X_v-Y_v)
        \int^{Y_v}_0 A dX'
        \int^{[E_\mu+\alpha(X_{grd}-X)]/r_M}_{E_\mu+\alpha(X_{grd}-X)} B
        \mathrm{exp} \left[
            -C\int^X_{X'} \frac{T_0(X''_v)}{X''_v} dX''
        \right]
        CdE,
\end{split}
\label{eq:Solved}
\end{equation}
where $[\frac{\delta Q_{\mu}}{\delta T_0(Y_v)}]_1$ corresponds to the first line of the result and $[\frac{\delta Q_{\mu}}{\delta T_0(Y_v)}]_2$ is for the second line. 
Using the same technique, the functional derivative of $P$ at $T_0(Y_v)$ is
\begin{equation}
\frac{\delta P}{\delta T_0(Y_v)}
    =-D\frac{1}{Y_v \cos\theta} \Theta(X_{grd}-Y_v)\Theta(Y_v-X) 
    \cdot\mathrm{exp} \left[-\int_X^{X_{grd}} 
             D \frac{T_0(\widetilde{X}_v)}{\widetilde{X}_v} d\widetilde{X}
        \right].
\label{eq:dPdT}
\end{equation}

Next is to integrate Eq.~\ref{eq:Solved} and \ref{eq:dPdT} over the slant depth $X$, muon energy $E_{\mu}$, and incident angle as in Eq.~\ref{eq:Deri1} and \ref{eq:Deri2}.
For the integration over X, the delta and step functions can be used to rewrite the variable names and integration limits.

With all the above results, the parameter $\alpha_T$ with complete contributions from both pions and the kaons is
\begin{equation}
    \alpha_T = \frac
    {\int_0^{X_{grd,v}} 
        \int \sin\theta d\theta d\phi \int_{E_{th}(\theta,\phi)}^\infty dE_{\mu} 
        \int_0^{X_{grd}(\theta,\phi)} 
        \frac{\delta (Q_\mu P)}{\delta T_0(Y_v)} dX
        T_0(Y_v) dY_v} 
    {\int \sin\theta d\theta d\phi \int_{E_{th}                 (\theta,\phi)}^\infty dE_{\mu}
        \int_0^{X_{grd}(\theta,\phi)} 
        (Q_{\pi\mu}+Q_{K\mu}) P dX},
\label{eq:finalAlpha}
\end{equation}
in which $\frac{\delta (Q_\mu P)}{\delta T_0(Y_v)}$ is 
\begin{equation}
    \frac{\delta (Q_\mu P)}{\delta T_0(Y_v)} = 
        [\frac{\delta Q_{\pi\mu}}{\delta T_0(Y_v)}
        +\frac{\delta Q_{K\mu}}{\delta T_0(Y_v)}]_1 P +
        [\frac{\delta Q_{\pi\mu}}{\delta T_0(Y_v)}
        +\frac{\delta Q_{K\mu}}{\delta T_0(Y_v)}]_2 P 
        +(Q_{\pi\mu}+Q_{K\mu}) \frac{\delta P}{\delta T_0(Y_v)}, 
\label{eq:finalWeight}
\end{equation}
where $Q_{\pi\mu}$, $Q_{K\mu}$, and $P$ are from Eq.~\ref{eq:MuonRate} and \ref{eq:MuonRateSupp}.
The weight reduces to
\begin{equation}
\begin{split}
W(Y_v) =&
\int \sin\theta d\theta d\phi \int_{E_{th}(\theta,\phi)}^\infty dE_{\mu} 
    \int_0^{X_{grd}(\theta,\phi)}  \\
    &\left\{
        [\frac{\delta Q_{\pi\mu}}{\delta T_0(Y_v)}
        +\frac{\delta Q_{K\mu}}{\delta T_0(Y_v)}]_1 P +
        [\frac{\delta Q_{\pi\mu}}{\delta T_0(Y_v)}
        +\frac{\delta Q_{K\mu}}{\delta T_0(Y_v)}]_2 P 
        +(Q_{\pi\mu}+Q_{K\mu}) \frac{\delta P}{\delta T_0(Y_v)} \right\} dX.
\end{split}
\label{eq:weight}
\end{equation}

At the end of this section, we decompose the above equation into three terms, namely the first local term $W_{\rm Q,loc}(Y_v)$, the second propagation term $W_{\rm Q,prop}(Y_v)$, and the third survival probability term $W_P(Y_v)$ 
\begin{equation}
    W(Y_v) = W_{\rm Q,loc}(Y_v) + W_{\rm Q,prop}(Y_v) +  W_{\rm P}(Y_v).
\label{eq:weightSum}
\end{equation}
The physical meaning of the local, propagation, and survival probability terms is given in Sec.~\ref{sec:solution}, for example, Sec.~\ref{sec:MuonRate}.
\end{widetext}

\begin{figure}[h]
  \centering  \includegraphics[width=0.4\textwidth]{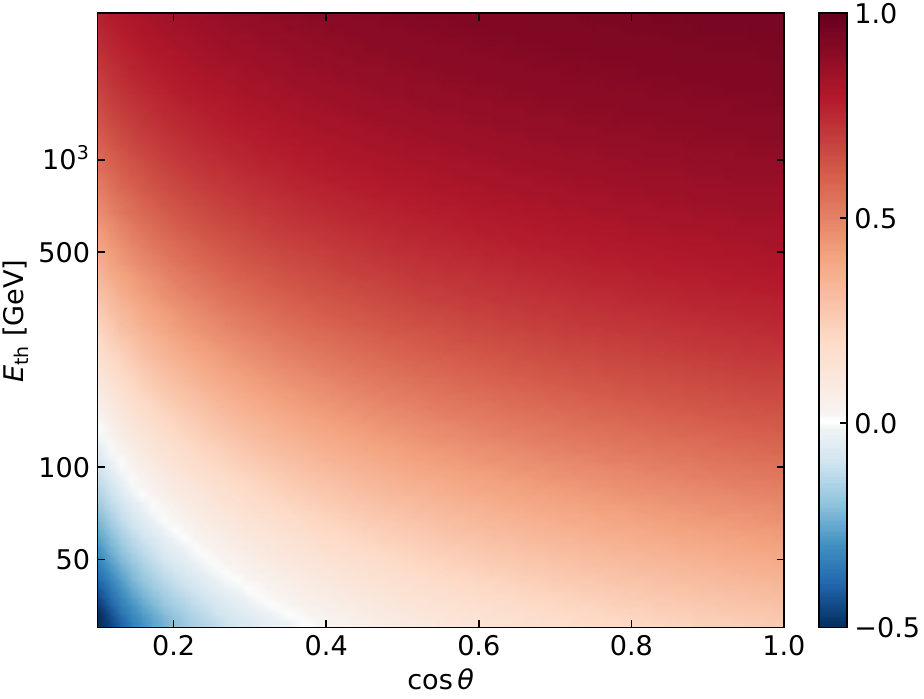}
  \caption{Theoretical result of the linear dependence parameter $\alpha_T(E_{th},\cos\theta)$ of Eq.~\ref{eq:finalAlpha} for specific $E_{th}$ and $\cos\theta$ over a wide range of $E_{th}$ and $\cos{\theta}$.}
  \label{fig:alpha_theo}
\end{figure}

\begin{figure}[h]
  \centering  \includegraphics[width=0.4\textwidth]{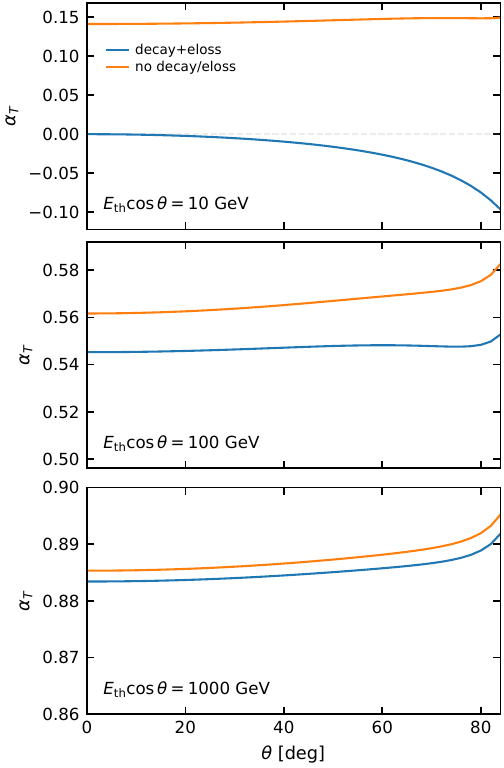}
  \caption{Linear dependence parameter $\alpha_T(E_{th},\cos\theta)$ (with or without the muon decay and energy loss contribution) as a function of $\theta$.
  Here $\alpha_T$ of Eq.~\ref{eq:finalAlpha} is calculated by scanning through all $E_{th}$ and $\cos\theta$ parameter pairs for three fixed $E_{th}\cos\theta$ values of 10 GeV, 100 GeV, and 1000 GeV. 
  The decay and energy-loss term in Eq.~\ref{eq:PElossDecay} is significant at low energy but negligible at higher energies. 
  The new theory does not guarantee that $\alpha_T$ is the same for the same $E_{th}\cos\theta$. 
  }
  \label{fig:alpha_fixed_ecos}
\end{figure}

Figure~\ref{fig:alpha_theo} shows the theoretical result of the linear dependence coefficient $\alpha_T(E_{th},\cos\theta)$ for a wide range of $E_{th}$ and $\cos{\theta}$, i.e.~the $\alpha_T$ result in Eq.~\ref{eq:finalAlpha} with specific and fixed $E_{th}$ and $\cos\theta$ (Eq.~\ref{eq:finalAlpha} without angular integration).
These formulas do not directly show dependence on the product $E_{th}\cos\theta$; that is, $\alpha_T(E_{th},\cos\theta)$ is not guaranteed to be the same for the constant $E_{th}\cos\theta$. 
Figure~\ref{fig:alpha_fixed_ecos} shows the linear dependence parameter $\alpha_T(E_{th},\cos\theta)$ as a function of $\theta$ for three constant $E_{th}\cos\theta$ values of 10 GeV, 100 GeV, and 1000 GeV.
This figure plots two cases: the full formula of Eq.~\ref{eq:finalAlpha} and one that assumes no energy loss and no muon decay.
The decay and energy-loss term in Eq.~\ref{eq:PElossDecay} is significant at low energy but negligible at higher energies. 
For $\theta>60^\circ$, $\alpha_T$ changes significantly.
This characteristic differs from previous literature~\cite{KPionRatio}.
Fortunately, for most underground muon detectors, most muons come from small angles (i.e., less than 60 degrees), where the change of $\alpha$ can be neglected.

\begin{figure}[h]
  \centering
  \includegraphics[width=0.4\textwidth]{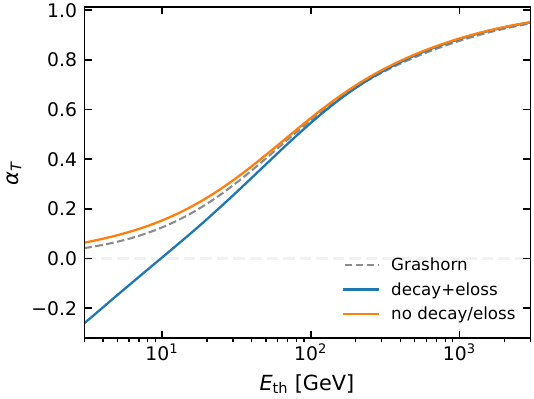}
  \caption{Coefficient $\alpha_T$ as a function of the energy threshold $E_{th}$ for the special case of $\theta=0$.
  The complete solution in Eq.~\ref{eq:finalAlpha} agrees with Ref.~\cite{KPionRatio} at high energy (1000 GeV) if assuming no muon energy loss or decay, but is slightly smaller at low energy (1 GeV); with realistic muon energy loss and decay, it differs significantly.
  }
  \label{fig:alpha_Eth}
\end{figure}

Figure~\ref{fig:alpha_Eth} shows how $\alpha_T(E_{th},\cos\theta)$ depends on the energy threshold $E_{th}$ in the special case $\theta=0$.
Eq.~\ref{eq:finalAlpha} without muon energy loss and decay agrees with Ref.~\cite{KPionRatio} at high energy (1000 GeV) but is slightly smaller at low energy (1 GeV); with realistic muon energy loss and decay, it differs significantly.

\begin{figure}[]
  \centering  \includegraphics[width=0.45\textwidth]{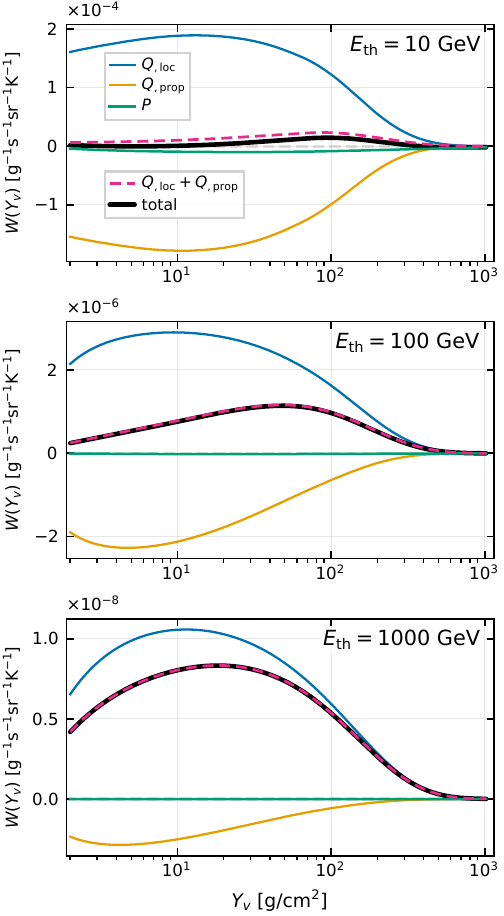}
  \caption{
  The three weight components in Eq.~\ref{eq:weightSum} versus depth $Y_v$ for three energies of 10 GeV, 100 GeV, and 1000 GeV. $\theta$ is set to zero. At high energies (100 and 1000 GeV), the contributions from $W_{\rm Q,loc}(Y_v) + W_{\rm Q,prop}(Y_v)$ dominate. At 10 GeV, the contribution from $W_{\rm P}(Y_v)$ becomes significant.}
  \label{fig:three_weight}
\end{figure}

Figure~\ref{fig:three_weight} shows the weights of the three components in Eq.~\ref{eq:weightSum} versus depth $Y_v$. At high energies (100 and 1000 GeV), the contributions from $W_{\rm Q,loc}(Y_v) + W_{\rm Q,prop}(Y_v)$ dominate. At 10 GeV, the contribution from $W_{\rm P}(Y_v)$ becomes significant.

\section{Numerical Test using MCEq}
\label{sec:MCEq}
\subsection{MCEq Config and Data Preparation}
\label{sec:config}

To test the theoretical picture developed above, we simulate the propagation of cosmic rays in the atmosphere using the Matrix Cascade Equations (MCEq)~\cite{MCEq_source}.
It is an open-source numerical toolkit for the coupled cascade equations of air showers, which can provide lepton fluxes at given atmospheric depths.
Compared with Monte Carlo programs such as CORSIKA, MCEq is much faster.
On a personal-computer CPU, a full simulation takes only a few seconds and yields lepton fluxes at different depths.

There is no randomness in the solution of MCEq. MCEq can only perform one-dimensional simulations, as if all secondary particles would propagate in the same direction as their parents.
This makes it impossible to simulate the transverse development of an air shower.
Because it solves for particle spectra, it cannot track individual particles in the way Monte Carlo codes do.
It can only distinguish the parent particle type, for instance, muons from the decay of $\pi$ or K mesons.

MCEq is highly customizable.
Built-in atmospheric models such as NRLMSISE-00 empirical model~\cite{atmos_n0} are available. 
It also supports custom atmospheric models; these require specifying $\rho(h)$ or $\rho(X)$, the air density at each altitude or depth. 
It can accommodate different incident zenith angles $\theta$, primary cosmic ray flux models, and particle-air interaction models.

In our MCEq-based study, the SIBYLL23E interaction model~\cite{Riehn:2024prp} is adopted.
For the primary cosmic ray flux, we adopt the Hillas Gaisser 2012 model (H3a) \cite{h3a}.
We also tested the Simple27 model, which includes only protons and follows a simple 2.7 power law, $N_{0}(E) = C_{0}E^{-(\gamma+1)}$, and the conclusions are consistent too.

The atmospheric data for the Daya Bay experiment are taken from the ERA5 database \cite{era5}.
ERA5 database is organized horizontally by latitude and longitude on a  0.25$^\circ$ by 0.25$^\circ$ grid. 
Therefore, the atmospheric data for the three halls in Daya Bay are the same.
In the vertical direction, the data are indexed by pressure, with 37 layers ranging from 1000 hPa (sea level) to 1 hPa (approximately 45-50 km altitude). 
Multiple meteorological variables, such as temperature and wind speed, are available hourly.

In ERA5, we obtain gravitational potential and temperature at each pressure level. Gravitational potential is the gravitational potential energy of a 1kg object so it is used to determine the absolute altitude $h$.
Temperature is used to calculate air density through the ideal gas law.
The data span January 1, 2011, to December 31, 2021, with one entry per day at 8:00 a.m., covering the full duration of the Daya Bay experiment. 

\begin{figure}[h]
  \centering
  \includegraphics[width=0.45\textwidth]{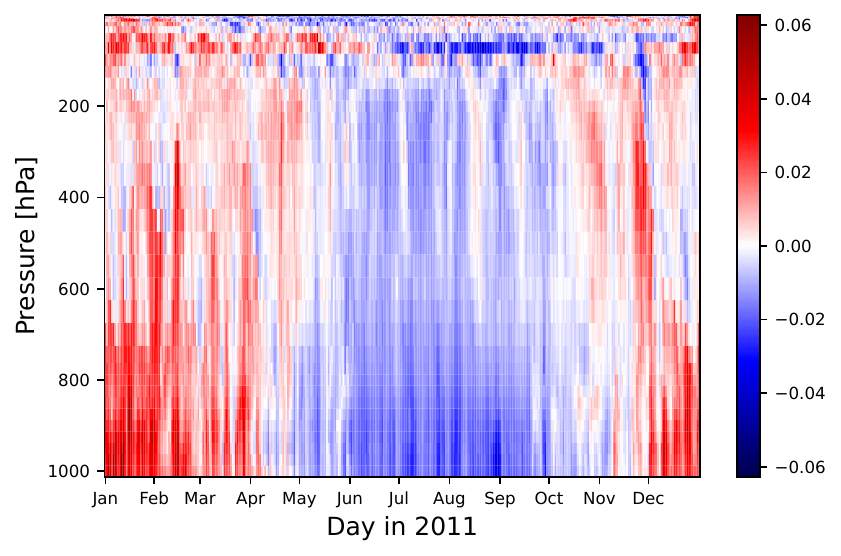}
  \caption{
  The variance of air density with respect to its average over time, $\frac{\rho(h,t) - \left \langle \rho(h) \right \rangle }{\left \langle \rho(h) \right \rangle }$, at various pressure levels above the Daya Bay site in 2011. 
  The air density and pressure levels are reported by ERA5~\cite{era5}.}
  \label{rela_rho}
\end{figure}

The variance of air density with respect to its average over time, $\frac{\rho(h,t) - \left \langle \rho(h) \right \rangle }{\left \langle \rho(h) \right \rangle } $, at various pressure levels above the Daya Bay site in 2011 is shown in Fig.~\ref{rela_rho}. 
Temperatures are higher in summer, so density is lower near the surface.

\begin{figure}[h]
  \centering
  \includegraphics[width=0.4\textwidth]{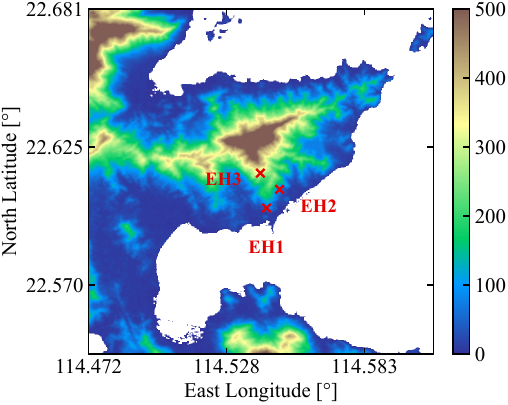}
  \caption{Terrain above the Daya Bay experiment and the positions of the three experimental halls.}
  \label{map}
\end{figure}

We obtained the terrain data around the Daya Bay experiment site from the database of Ref.~\cite{geo_data}. 
The terrain of the mountain above the Daya Bay experiment is reconstructed using two-dimensional linear interpolation.
Fig.~\ref {map} shows the topographic map and detector positions.
Considering the information provided in Ref.~\cite{DYB-TAUP2025, DYB_muon_system, DYB_2,DYB2007},
the vertical overburdens are 93, 99, and 324 meters for Daya Bay EH1, EH2, and EH3, respectively.

For rock, we adopt the threshold-energy expression in Eq.~\ref{eq:e_loss} from Gassier~\cite{Gaisser}. 
For the muon energy loss in the atmosphere, we use the empirical formula built into the MCEq,
\begin{equation}
  E_{th} = \epsilon (e^{X_{rock}/\xi} -1)
\label{eq:e_loss}
\end{equation}
with $\xi = 2.5\times 10^{5}~\rm{g/cm^{2}}$ and $\epsilon =\alpha\xi=500~\rm{GeV}$.
Here $\alpha$ is an approximate constant energy loss factor $\alpha=2~\rm{MeV/(g/cm^{2})}$. We adopt a rock density of $2.6~\rm{g/cm^{3}}$ based on field exploration~\cite{DYB2007}.

\subsection{Daya Bay underground muon rate prediction}
\label{rate_prediction}

The algorithm for simulating the underground muon rate is as follows.
We start with the H3a model.
With the air density $\rho(X, T)$ as input to MCEq, the surface muon spectrum $\phi_{\mu}(E_{\mu},\theta, \phi)$ is obtained.
The threshold energy $E_{\rm th}(\theta,\phi)$ is determined according to the terrain data using Eq.~\ref{eq:e_loss}.
Then we integrate the spectrum over energy from the threshold energy at each $\theta$ and $\phi$, multiplied by the corresponding solid angle, and sum all contributions to obtain the total muon rate. 
The difference in maximum atmospheric depth $X_{grd}$ caused by terrain is ignored.
A unified maximum (vertical) depth $X_{grd}=1033~\rm g/cm^{2}$ is adopted. 

\begin{figure}[h]
  \centering
  \includegraphics[width=0.45\textwidth]{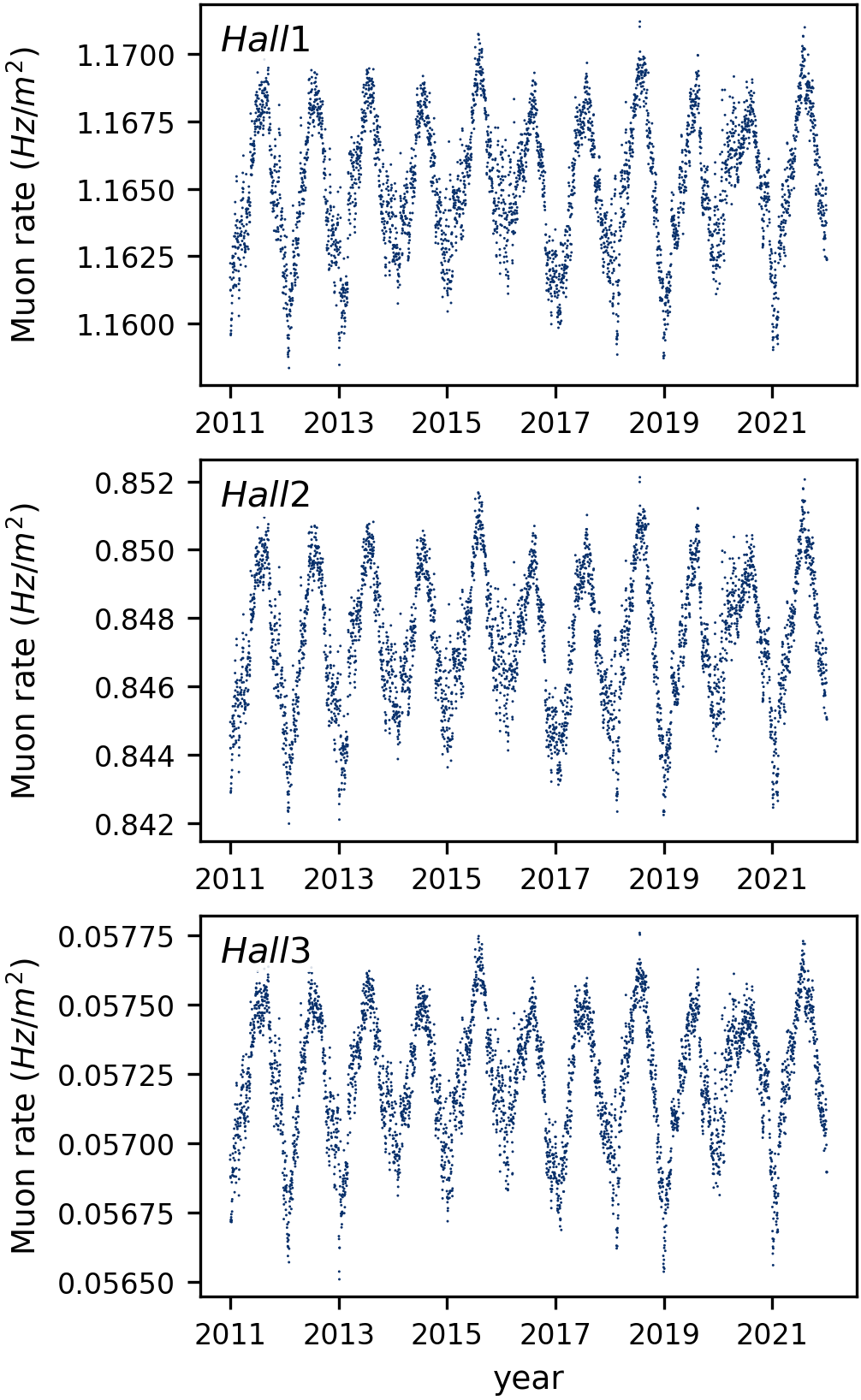}
  \caption{Simulated underground muon rate variation versus time at Daya Bay Experiment. Seasonal modulation can be found.}
  \label{muon_rate_pic}
\end{figure}

Table~\ref{muon_rate_table} shows the predicted underground muon rates in the three experimental halls and compares them with the experimental results. Good agreement is found.
Figure~\ref{muon_rate_pic} shows the variation of the underground muon rate over the 11 years of Daya Bay data-taking (from January 1, 2011, to December 31, 2021, at 8 a.m. daily), revealing its seasonal modulation.
Note that the absolute muon rate would differ if choosing a different hadronic interaction model in MCEq.
\begin{table}[h]
\centering
\begin{tabular}{c|cc}
      & MCEq Simulation [$\rm Hz/m^{2}$]     & Experiment [$\rm Hz/m^{2}$]~\cite{DYB_muon_system} \\
  \hline
  EH1                   & 1.165                        & 1.16$\pm$0.11                 \\
  EH2                   & 0.847                        & 0.86$\pm$0.09                 \\
  EH3                   & 0.057                        & 0.054$\pm$0.006                 \\
\end{tabular}
\caption{Predicted and measured underground muon rates at Daya Bay Experiment.}
\label{muon_rate_table}
\end{table}

\subsection{Repeat of the nonlinear effect at Daya Bay}
\label{sec:nonlinear_repeat}

\begin{figure*}[htb]
  \centering
  \includegraphics[width=0.9\textwidth]{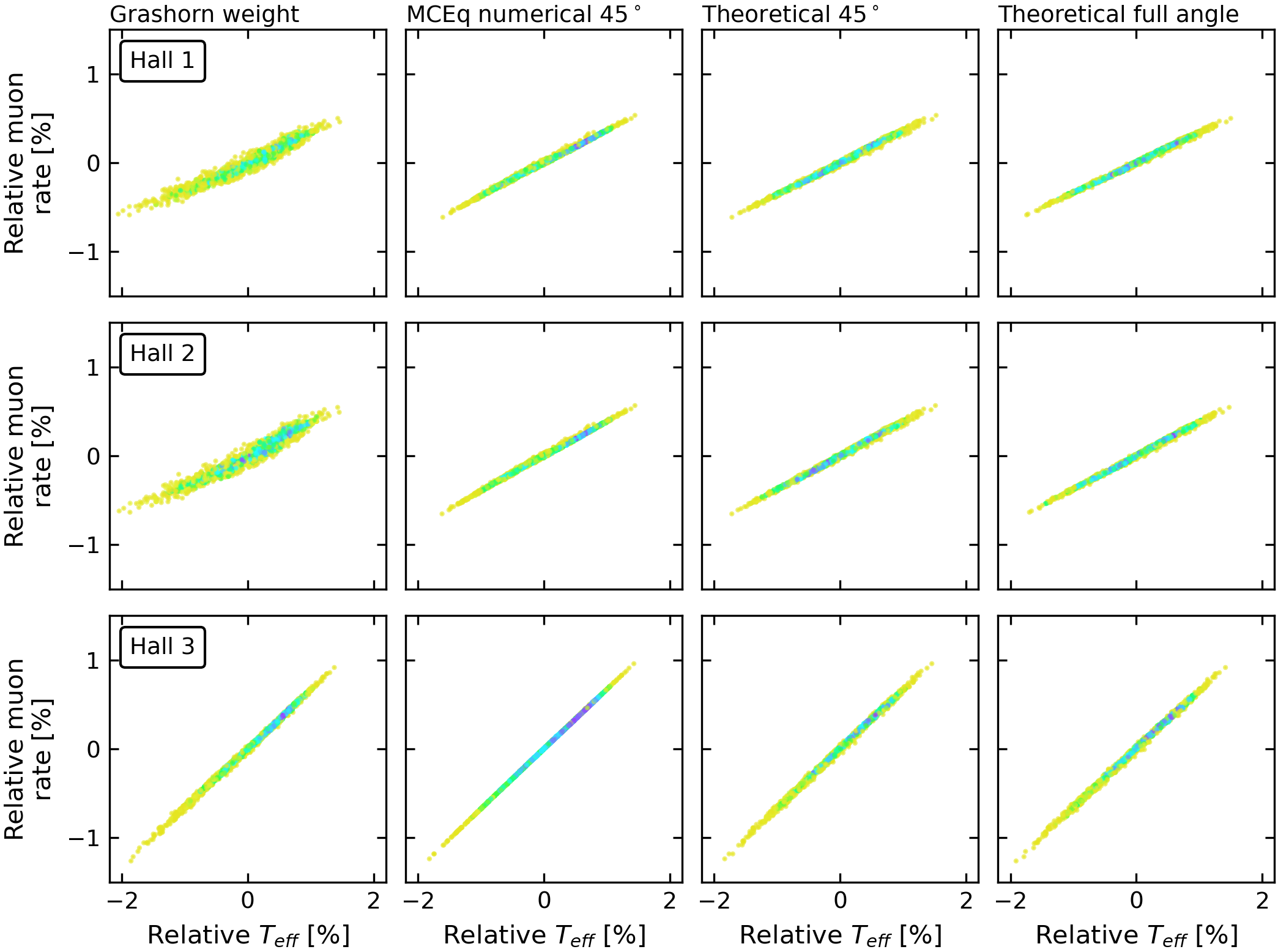}
  \caption{Correlation between the relative change of underground muon rate and the relative change of effective temperature. 
  The three rows correspond to the three experimental halls at Daya Bay.
  The muon rates are simulation results with MCEq, using the real temperature profile from ERA5.
  In the first column, 
  the muon rate results are for the total muon rate of all angles for each hall, and 
  the effective temperature is calculated using the weights of Grashorn~\cite{KPionRatio}.
  In the second column, the muon rate results are for a special zenith angle of $\theta=45^\circ$ with $E_{th}\cos\theta$ equal to the average of the corresponding hall, 
  but the temperature weights are extracted from MCEq calculation.
  In the third column, the muon rate results are the same as in column two, but the temperature weights are based on Eq.~\ref{eq:weightSum}, which recovers a linear dependency with the new theory.
  In the last column, the muon rate results are again shown for all angles at each site, as in column one, but the temperature weights are based on Eq.~\ref{eq:weightSum}, where a linear response can be found.
  The MCEq weights best explain its own simulated rate effect at high energy of Hall 3 because there are no calculation errors and no significant muon decay and energy loss effects.
  }
  \label{fig:nonlinear_rrrt}
\end{figure*}

We reproduced the nonlinear effect observed at Daya Bay, as shown in the three subfigures in the first column of Fig.~\ref{fig:nonlinear_rrrt}.
It shows how relative change in the underground muon rate correlates with the relative change in the effective temperature. The effective temperature is weighted according to Grashorn~\cite{KPionRatio}, including the kaon contribution.
The Daya Bay nonlinearity is reproduced for EH1, EH2, and EH3.
The temperature coefficient $\alpha_{T}$ increases with overburden.

The average values of energy thresholds multiplied $\cos\theta$, $\left \langle E_{th}cos \theta \right \rangle$, are 44, 51, and 187 GeV for the three halls, respectively.
Note that this differs from the original Daya Bay results of 37, 41, and 143 GeV~\cite{DYB-TAUP2025}, because 
their $E_{th}$ at a zenith angle $\theta$ is the minimum over all azimuthal angles $2\pi$. This calculation underestimated the overall average.

\subsection{Linear dependency recovered}
\label{sec:recoverd}
We've tested if the linearity can be recovered. 
The weight used to calculate $T_{\rm eff}$ and $\Delta T_{\rm eff}$ can be obtained in two ways.
One is to use MCEq calculation, and the other way is to use pure theoretical prediction in Sec.~\ref{sec:weight}. The MCEq configuration and simulated data are the same as Sec.~\ref{sec:nonlinear_repeat}. The calculation process and results are presented below.

\subsubsection{Numerical method to calculate $\frac{\delta Q_{\mu}}{\delta T(Y)}$}
\label{sec:Numerical}

To obtain the numerical values of the weight $W$ and the temperature coefficient $\alpha_T$, we use MCEq to directly calculate muon spectra at each depth.
The algorithm is as follows.

The key step is to calculate the functional derivative of the production spectrum $\frac{\delta Q_{\mu}}{\delta T_0(Y_v)}$.
Here we ignore the contribution from muon energy loss and decay because it is not significant in the Daya Bay energy range; otherwise, the calculation would be much more complicated.
For simplicity, we use a generic functional $J[f(x_{k})]$ to explain the discrete calculation procedure. The functional derivative of $\frac{\delta J}{\delta f(x)}$ with a functional change of $\eta(x)$ should satisfy the following equation:
\begin{equation}
    \int \frac{\delta J}{\delta f(x)} \eta (x) dx = \lim_{h=0}\frac{J[f(x)+h\eta(x)]-J[f(x)]}{h}.
\end{equation}

For a discrete case, the parameter space is separated into many pieces, and each is labeled with an index $i$; assuming $h$ is small, it becomes 
\begin{equation}
    \sum_{i}^{}  \frac{\delta J}{\delta f(x_{i})} \eta(x_{i}) \Delta x_{i} =  \frac{J[f(x_{i})+h\eta(x_{i})]-J[f(x_{i})]}{h}.
\label{eq:dis_def}
\end{equation}
For the discrete case, the function $f(x_{i})=v$ means the function value is $v$ in a $\Delta x_{i}$-wide bin around $x_{i}$ and all these bins cover the full parameter space without omission or overlap.

Let us evaluate the functional derivative at $x_k$.
A perturbation is only introduced to $\eta$ at $x_k$ as
\begin{align}
\left\{
\begin{aligned}
\eta(x_i) \ne 0,~{\rm for}~i= k \\ 
\eta(x_i) = 0,  ~{\rm for}~i\ne k \\
\end{aligned}
\right. \\
{\rm for~} i=1,2,3\dots \nonumber
\end{align}
In this way, Eq.~\ref{eq:dis_def} can be simplified into 
\begin{equation}
 \frac{\delta J}{\delta f(x_{k})} = \frac{J[f(x_{i})+h\eta(x_{i})] - J[f(x_{i})]}{h\Delta x_{k}\eta(x_k)} .
\label{eq:num_functional_derivative}
\end{equation}

The $\Delta x_{k}$ in the denominator makes it consistent with the analytical result and ensures the dimensions are correct. 
Notice that the dimension of the functional derivative is not what it appears to be. The dimension of $\frac{\delta J}{\delta f(x)}$ turns out to be $[\frac{J}{f \cdot x}]$.

In the actual calculation, changes of $J\to Q_\mu$, $f \to T_0$, and $\eta \to \Delta T$ should be made.
The production spectrum of muons $Q_{\mu}$ is calculated by the difference of muon spectra in adjacent layers. 
Because the division of atmospheric depth bins is logarithmically uniform in MCEq, we set the $Q_{\mu}$ as below:
\begin{equation}
\begin{split}
  Q_{\mu}&(\sqrt{X_{i}X_{i+1}} ,E_{\mu})=Q_{\mu}(X_{i}\to X_{i+1} ,E_{\mu})\\
  &=\frac{1}{X_{i+1}- X_{i}}(\phi_{\mu}(X_{i+1},E_{\mu})-\phi_{\mu}(X_{i},E_{\mu})) .
\label{eq:muon_gen}
\end{split}
\end{equation}

Finally, the functional derivative of Eq.~\ref{eq:num_functional_derivative} is calculated 
using the MCEq output of Eq.~\ref{eq:muon_gen} at each depth $X$, and weight is calculated according to the discussion in Sec.~\ref{sec:weight}. 
With the new weight, the $T_{\rm eff}$ and $\Delta T_{\rm eff}$ are calculated for each temperature data point.
The new two-dimensional plots of relative muon rate changes versus the relative temperature changes for Daya Bay three halls are plotted in column two of Fig.~\ref{fig:nonlinear_rrrt}, in which we choose $\theta=45^\circ$ for demonstration
with $E_{th}\cos\theta$ equal to the average of that hall.
Linearity can be recovered with the correct weight.

\subsubsection{Theoretical weight}

For the real test goal, the effective temperature $T_{\rm eff}$ and the variation of $\Delta T_{\rm eff}$ are directly calculated purely with the theoretical prediction.

Column three of Fig.~\ref{fig:nonlinear_rrrt} first
shows the result at $\theta=45^\circ$, where, except for the weight, the rest of the configurations match the second column. 
The muon rates are MCEq simulation results with ERA5 temperature inputs. 
The effective temperature is calculated using the weights from the prediction of Eq.~\ref{eq:weightSum}.
The $E_{th}\cos\theta$ equals to the average of that hall.
The last column of Fig.~\ref{fig:nonlinear_rrrt} shows the results for the full solid-angle calculation of the Daya Bay halls.
Linearity is recovered.
  
\section{Summary}
\label{sec:summary}
In this work, we revisit the atmospheric temperature dependence of underground muon flux and investigate the nonlinear modulation effect observed in the Daya Bay Neutrino Experiment. We show that the conventional meson and muon spectrum solutions commonly adopted in the literature are approximate solutions. They are obtained under simplified atmospheric assumptions and do not fully account for the influence of non-local temperature variations. We use an improved solution of the atmospheric cascade equations that preserves the full temperature history along meson and muon trajectories and naturally connects the low- and high-energy regimes. Based on the functional dependence of the underground muon rate on the atmospheric temperature profile, new definitions of the effective temperature, its variation, and the temperature correlation coefficient are introduced. The MCEq-predicted nonlinear correlation between the underground muon rate and effective temperature is restored to a linear dependence by applying the functional derivative method. 


\section*{Acknowledgments}
This work is supported in part by
the National Natural Science Foundation of China (No.~12141503),
the Ministry of Science and Technology of China (No.~2022YFA1604704),
the Key Laboratory of Particle \& Radiation Imaging (Tsinghua University),
the CAS Center for Excellence in Particle Physics (CCEPP), and the College Student Innovation Training Program. We also thank Mr.~Peiyang Liu for the calculation cross-check.

\bibliographystyle{unsrt}
\bibliography{MuonSeasonal}

@article{Barrett,
    author = "Barrett, Paul H. and others",
    title = "{Interpretation of Cosmic-Ray Measurements Far Underground}",
    journal = "Rev. Mod. Phys.",
    volume = "24",
    number = "3",
    pages = "133",
    year = "1952"
}

@article{Sherman,
    author = "Sherman, Noah",
    title = "{Atmospheric Temperature Effect for $\mu$ Mesons Observed at a Depth of 846 m.w.e.}",
    journal = "Phys. Rev.",
    volume = "93",
    number = "1",
    pages = "208",
    year = "1954"
}

@inproceedings{Poatina,
    author = "Humble, J. E. and others",
    title = "{Variations in atmospheric coefficients for underground cosmic ray detectors}",
    booktitle = "{16th International Cosmic Ray Conference}",
    volume = "4",
    pages = "258",
    year = "1979"
}

@inproceedings{Baksan,
    author = "Andreev, Yu. M. and others",
    title = "{Season and daily variations of the intensity of muons with $E_{\mu} \geq 220$~GeV}",
    booktitle = "{21st International Cosmic Ray Conference}",
    volume = "7",
    pages = "88",
    year = "1990"
}

@article{MACRO,
    author = "Ambrosio, M. and others",
    title = "{Seasonal variations in the underground muon intensity as seen by MACRO}",
    journal = "Astropart. Phys.",
    volume = "7",
    number = "1-2",
    pages = "109",
    year = "1997"
}

@inproceedings{AMANDA,
    author = "Bouchta, A.",
    title = "{Seasonal variation of the muon flux seen by AMANDA}",
    booktitle = "{26th International Cosmic Ray Conference}",
    pages = "108",
    year = "1999"
}

@inproceedings{IceCube,
    author = "Desiati, P. and others",
    title = "{Seasonal Variations of High Energy Cosmic Ray Muons Observed by the IceCube Observatory as a Probe of Kaon/Pion Ratio}",
    booktitle = "{32nd International Cosmic Ray Conference}",
    volume = "1",
    pages = "78",
    year = "2011"
}

@article{MINOS,
    author = "Adamson, P. and others",
    title = "{Observation of muon intensity variations by season with the MINOS near detector}",
    journal = "Phys. Rev. D",
    volume = "90",
    number = "1",
    pages = "012010",
    year = "2014"
}

@article{Borexino,
    author = "Bellini, G. and others",
    title = "{Cosmic-muon flux and annual modulation in Borexino at 3800 m water-equivalent depth}",
    journal = "J. Cosmol. Astropart. Phys.",
    volume = "05",
    pages = "015",
    year = "2012"
}

@article{GERDA,
    author = "Agostini, M. and others",
    title = "{Flux modulations seen by the muon veto of the GERDA experiment}",
    journal = "Astropart. Phys.",
    volume = "84",
    pages = "29",
    year = "2016"
}

@article{DoubleChooz,
    author = "Abrah{\~a}o, T. and others",
    title = "{Cosmic-muon characterization and annual modulation measurement with Double Chooz detectors}",
    journal = "J. Cosmol. Astropart. Phys.",
    volume = "02",
    pages = "017",
    year = "2017"
}

@book{Gaisser,
    author = "Gaisser, Thomas K. and Engel, Ralph and Resconi, Elisa",
    title = "{Cosmic Rays and Particle Physics}",
    publisher = "Cambridge University Press",
    year = "2016"
}

@article{KPionRatio,
    author = "Grashorn, E. W. and others",
    title = "{The atmospheric charged kaon/pion ratio using seasonal variation methods}",
    journal = "Astropart. Phys.",
    volume = "33",
    number = "3",
    pages = "140",
    year = "2010"
}

@inproceedings{DYB-ICRC2025,
    author = "Dugas, K.",
    title = "{Towards a Measurement of the Seasonal Variation of the Muon Flux Using the Full Daya Bay Dataset}",
    booktitle = "{39th International Cosmic Ray Conference (ICRC 2025)}",
    year = "2025"
}

@inproceedings{DYB-TAUP2025,
    author = "Ma, Bangzheng",
    title = "{Seasonal Variation of Muon Flux in Daya Bay Using the Full Data Set}",
    booktitle = "{19th International Conference on Topics in Astroparticle and Underground Physics (TAUP 2025)}",
    year = "2025"
}

@article{Bangzheng,
    author = "Ma, Bangzheng and others",
    title = "{Biases in the Determination of Correlations Between Underground Muon Flux and Atmospheric Temperature}",
    journal = "arXiv:2604.06704",
    year = "2026"
}

@article{MCEq_source,
    author = "Fedynitch, Anatoli and others",
    title = "{Calculation of conventional and prompt lepton fluxes at very high energy}",
    journal = "EPJ Web Conf.",
    volume = "99",
    pages = "08001",
    year = "2015"
}

@article{atmos_n0,
    author = "Picone, J. M. and others",
    title = "{NRLMSISE-00 empirical model of the atmosphere: Statistical comparisons and scientific issues}",
    journal = "J. Geophys. Res. Space Phys.",
    volume = "107",
    number = "A12",
    pages = "1468",
    year = "2002"
}

@article{Riehn:2024prp,
    author = "Riehn, Felix and Fedynitch, Anatoli and Engel, Ralph",
    title = "{Sibyll★}",
    journal = "Astropart. Phys.",
    volume = "160",
    pages = "102964",
    year = "2024"
}

@article{h3a,
    author = "Gaisser, Thomas K.",
    title = "{Spectrum of cosmic-ray nucleons, kaon production, and the atmospheric muon charge ratio}",
    journal = "Astropart. Phys.",
    volume = "35",
    number = "12",
    pages = "801",
    year = "2012"
}

@article{era5,
    author = "Hersbach, Hans and others",
    title = "{The ERA5 global reanalysis}",
    journal = "Q. J. R. Meteorol. Soc.",
    volume = "146",
    number = "730",
    pages = "1999",
    year = "2020"
}

@misc{geo_data,
    author = {{Computer Network Information Center, Chinese Academy of Sciences}},
    title = {Geospatial Data Cloud site},
    year = {2020}
}

@article{DYB_muon_system,
    author = "An, F. P. and others",
    title = "{The muon system of the Daya Bay Reactor antineutrino experiment}",
    journal = "Nucl. Instrum. Methods Phys. Res. A",
    volume = "773",
    pages = "8",
    year = "2015"
}

@article{DYB_2,
    author = "An, F. P. and others",
    title = "{Measurement of electron antineutrino oscillation based on 1230 days of operation of the Daya Bay experiment}",
    journal = "Phys. Rev. D",
    volume = "95",
    number = "7",
    pages = "072006",
    year = "2017"
}

@article{DYB2007,
    author = "Guo, Xinheng and others",
    title = "{A Precision Measurement of the Neutrino Mixing Angle $\theta_{13}$ using Reactor Antineutrinos at Daya Bay}",
    journal = "arXiv:hep-ex/0701029",
    year = "2007"
}

@article{Previous,
title = {Corrections for temperature effect for ground-based muon hodoscopes},
journal = {Astroparticle Physics},
volume = {34},
number = {6},
pages = {401-411},
year = {2011},
issn = {0927-6505},
doi = {https://doi.org/10.1016/j.astropartphys.2010.10.013},
url = {https://www.sciencedirect.com/science/article/pii/S0927650510002100},
author = {A.N. Dmitrieva and R.P. Kokoulin and A.A. Petrukhin and D.A. Timashkov},
}

\end{document}